\documentclass{ab}

\usepackage{graphicx}
\usepackage{natbib}

\def\degr{\hbox{$^\circ$}}
\def\arcmin{\hbox{$^\prime$}}

\begin{document}

\title{Technique of Polarimetric Observations of Faint Objects
\\ at the 6-m BTA Telescope}

\author{
{V.~L.}~{Afanasiev}$^1$, {V.~R.}~{Amirkhanyan}$^2$ }

\institute{ $^1$Special Astrophysical Observatory, Russian Academy
of Sciences, Nizhnij Arkhyz, 369167, Russia\\
$^2$Sternberg State Astronomical Institute, Moscow State
University, Moscow, 119991 Russia}

\titlerunning{POLARIMETRIC OBSERVATIONS OF AT THE BTA}
\authorrunning{AFANASIEV \& AMIRKHANYAN}

\date{June 13, 2012/September 20, 2012}
\offprints{Viktor Afanasiev \email{vafan@sao.ru}}

\abstract{ We describe the technique of spectropolarimetric
observations allowing for the measurements of the Stokes
parameters in one of the observational modes of the SCORPIO focal
reducer of the 6-m BTA telescope of the SAO RAS. The
characteristics of the instrument in the spectropolarimetric mode
of observations are given. We present the algorithm of
observational data reduction. The capabilities of the SCORPIO
spectropolarimetric mode are demonstrated on the examples of
observations of various astronomical objects.}

\maketitle

\section{INTRODUCTION}

Classical methods of polarimetry and spectropolarimetry, used for
observing faint objects are based on the application of
double-beam schemes and photon counters, enabling fast switching
of the positions of phase elements and thus suppressing the
atmospheric scintillation, which is a major obstacle in  the
polarimetric observations~\citep{piirola:Afanasiev_n,
shakh:Afanasiev_n}.

Modern trends in the techniques of observations of faint objects,
applied at large telescopes are based on the use of focal reducers
and slow panoramic detectors (CCDs). Specifically, over the recent
years the focal reducers are being complemented with the
polarimetric observation modes ~\citep{apen:Afanasiev_n,
foca:Afanasiev_n, kawa:Afanasiev_n}, which allow for the
measurements of polarization of faint objects with an intermediate
and low spectral resolution. The spectropolarimetric data are
important for the study of the physics of such objects as active
galactic nuclei, supernovae, strongly magnetized white dwarfs,
etc. and in many cases permit to obtain the information on the
nature of the detected radiation, to study the geometric and
magnetic characteristics of the objects.

The SCORPIO focal reducer is quite successfully used at the 6-m
BTA telescope of the Special Astrophysical Observatory of the
Russian Academy of Sciences (SAO RAS) since 2000. The SCORPIO
stands for the Spectral Camera with Optical Reducer for
Photometric and Interferometric Observations. The instrument is
installed at the primary focus of the telescope and features a
variety of observational modes: field photometry in the wide,
intermediate and narrow-band filters, panoramic spectroscopy with
the Fabry-Perot interferometer, long-slit spectroscopy, multislit
spectroscopy and spectropolarimetry~\citep{afan1:Afanasiev_n}. In
the latter case, a rotating Savar plate  is used as a polarization
analyzer, installed after a set of slits (diaphragms). A
description of the technique of observations and data reduction
can be found in~\citep{afan2:Afanasiev_n}. This instrument was
used to measure the linear and circular polarization in the
spectra of faint quasars and stars~\citep{afan3:Afanasiev_n,
afan4:Afanasiev_n}. The main setbacks of the observational
technique using the Savar plate are different chromatic
aberrations for the ordinary and extraordinary rays, leading to
the differences in the spectral line contours, and a small height
of the slit---in the observational mode, described
in~\citep{afan2:Afanasiev_n} it amounts to only $9''$, what
renders exceedingly difficult the subtraction of the sky
background in extended objects. This fact, combined with many
others, has prompted the development of a next generation focal
reducer for the BTA, called the  SCORPIO-2. The instrument is
designed for the application of a large-format CCD and possesses
significantly advanced capabilities, as compared to the SCORPIO-1,
currently operated at the BTA (as to the number of filters and
gratings used, a fast input into the Fabry-Perot interferometer
beam, the integration of the IFU mode, etc.).
 We report here the design features of the polarimetric observations mode in the new
SCORPIO-2 focal reducer of the 6-m BTA telescope. The technique of
observations and the algorithms of reduction of the obtained data
are also described.

\begin{figure*}

\includegraphics[scale=0.7]{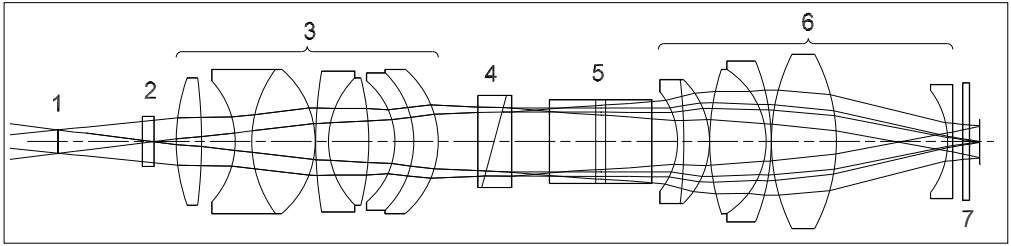}
\caption{The optical scheme of the polarimetric mode of the next
generation focal reducer SCORPIO-2: 1---slit; 2---the phase plate;
3---the collimator, 4---the Wollaston prism; 5---the grism;
6---the spectrograph camera; 7---the entrance window of the CCD
cryostat.} \label{layout:Afanasiev_n}
\end{figure*}

\section{THE SCHEME OF POLARIZATION MEASUREMENTS}

For the measurements of polarization of the registered signal we
have selected an optical scheme, consisting of
 rotating phase plates and a fixed
polarization analyzer. This design allows to unify the scheme in
the transition between the measurements of linear and circular
polarizations: the phase plate, shifting the phase by $\lambda/2$
is replaced by a  plate with a shift of $\lambda/4$.
Figure~\ref{layout:Afanasiev_n} presents the optical scheme of the
SCORPIO in the configuration of the polarimetric and
spectropolarimetric  mode of observations. The transition to the
polarimetric measurements is carried out by the insertion into the
optical path of the instrument of a prism and a precision unit,
containing the phase plates and a dichroic analyzer (a polaroid),
designed to measure the linear polarization of starlike and
extended objects in the field of view with the diameter of
6$\arcmin$.

The polarization analyzers used are the two Wollaston prisms,
designated WOLL-1 and WOLL-2 and a dichroic polarization filter
(POLAROID).

WOLL-1 is a Wollaston prism made of the calcite block crystal,
which has an octahedral shape sized 55$\times$55~mm and is 17~mm
thick. The angle of divergence of the ordinary and extraordinary
rays is $5\degr$, what corresponds to the operating slit height of
about 2\arcmin\ on the celestial sphere.

WOLL-2 is a composite optical element consisting of four Wollaston
prisms with a pairwise orientation of the optical axes, namely,
$0\degr$, $90\degr$ and $45\degr$, $135\degr$. All the prisms are
manufactured of the calcite block crystal, which has a tetrahedral
shape, sized 25$\times$25~mm and is 10~mm thick. The angle of
divergence of the ordinary and extraordinary rays amounts to
$2\fdg5$, and the operating height of the slit makes up about
1\arcmin\ on the celestial sphere.

The polarization analyzers are mounted on the rotating turret in
order to be inserted into the optical path of the instrument,
which yields the precision of the unit of at least 0\farcm1. The
precision block contains three phase polarization elements: the
achromatic $\lambda/2$ and $\lambda/4$ plates  30~mm  in diameter,
and a dichroic polaroid 70~mm in diameter, transmitting the
radiation in only one polarization plane. All the three elements
are installed in rotating frames and are inserted into the beam
via a linear displacement. The shifting mechanism has four fixed
positions: $\lambda/2$, $\lambda/4$, Polaroid and Hole, and is
mounted on a flat plate. The unit is also equipped
with two ten-slot filter 
turrets to insert the interference filters into the beam. The unit
contains the following elements of control and monitoring.
\begin{list}{}{
\setlength\leftmargin{2mm} \setlength\topsep{1mm}
\setlength\parsep{-0.5mm} \setlength\itemsep{2mm} }

\item 1)  A stepper motor rotating the screw spindle of the
carriage for changing the phase plate, with zero-point and
position fixing.  The repeatability of position setting is at
least $10~\mu$m.

\item 2)  A stepper motor setting the angle of rotation of phase
elements. Each phase element is installed at fixed positions by
the angle of rotation  ($\lambda/2$ plate at $0\degr$, $22\fdg5$,
$45\degr$, $67\fdg5$, $\lambda/4$ plate at $0\degr$ and $90\degr$,
the polarization filter at $0\degr$, $60\degr$, $120\degr$) with
mechanical locking. The repeatability of setting the angle of
rotation is better than $0\fdg02$. The indication of each position
by the angle of rotation is effected by the Hall sensors.

\item 3)  A stepper motor of the mechanism rotating the filter
turrets and fixing the positions; the Hall sensors are used to
indicate each of the ten positions.

\end{list}

The unit is mounted on a flat plate with one side housing the
mechanism moving the phase elements, while the other side houses
the filter turrets. The element control is done via a
microprocessor.

The observations in the WOLL-1 and POLAROID modes are conducted in
the packet mode, which allows making a sequence of exposures with
different settings of the angle of rotation of the phase plate or
polarization filter. The menu setting up the switching sequences
is called from the user interface, controlling the SCORPIO-2
spectrograph. The number of cycles is not limited by anything
except the duration of night.

Setting the {``lambda/2''} flag in the user interface, the
$\lambda/2$ phase plate is inserted in the convergent beam, which
is successively  set in each cycle in four positions by the angle:
$0\degr$, $45\degr$, $22\fdg5$ and $67\fdg5$. Setting the
{``lambda/4''} flag allows to successively rotate the inserted
$\lambda/4$ plate in two positions by the angle of 0\degr and
$90\degr$. Finally, setting the {``Polaroid''} flag, the dichroic
analyzer is positioned in three angles, $+60\degr$, $0\degr$, and
$-60\degr$. Setting the {``lambda/2\,+\,lambda/4''} flag  allows
for a full cycle of switchings required for the measurement of all
four Stokes parameters. The duration of the exposure, its type
({\tt obj}, {\tt flat}, {\tt neon} and {\tt eta}), as well as the
parameters of the CCD readout (the readout area and binning, the
GAIN and readout RATE) are set in the SCORPIO-2 exposure control
interface menu. The number of repeated exposures in the WOLL-2
mode observations is set in the same menu. Performing the
observations in the WOLL-2 mode, the zero polarization standard
stars have to be observed during the night to calibrate the
sensitivity of all four polarization channels.

\section{CALCULATION OF THE STOKES PARAMETERS}

It is well known that the parameters of a polarized
quasi-monochromatic wave are adequately described by the
components of the Stokes four-vector (see, e.g., the survey by
Rosenberg~\citep{rozen:Afanasiev_n}). The Stokes $I$,~$Q$,~$U$,
and~$V$ parameters possess the dimension of intensity or
electromagnetic radiation flux and correspond to the radiation
intensity differences with various types of polarization. Namely,
$I$ represents the wave intensity, $Q$ and $U$ characterize the
linear polarization, and $V$ specifies the circular polarization
of electromagnetic radiation. A complete definition of the Stokes
parameters and their relation to the parameters of the
corresponding fluctuations of the electromagnetic field can be
found in~\mbox{\citep{serk:Afanasiev_n, chandr:Afanasiev_n,
shurk:Afanasiev_n}}. In the schemes based on  fast switching of
phase elements the recorded intensity is modulated, and the
measured signal is, in essence, a linear combination of the
products of parameters by the values of harmonic functions for
different angles. The measurement of the Stokes parameters is
reduced to solving a system of linear equations. At that, if the
angles of phase plates are fairly well broken up within the
semi-circle, and this procedure is iterated a great many times,
the accuracy of measurement of the polarization parameters is
limited only by the count
statistics~\citep{tinbergen:Afanasiev_n}.

Observing the faint objects, rather long exposures have to be
taken, and there is no possibility of frequent and detailed
switching the phase elements by the angle. It should also be borne
in mind that in real observational conditions the resultant signal
from the object and the background sky is always recorded.
Furthermore, background radiation is often linearly polarized,
especially at large moon phases or in the dusk, sometimes this
polarization can reach up to $20$--$30\%$. To obtain the
undistorted information on the polarization of radiation of the
studied object it is necessary to subtract the sky background,
obtained at similar positions of the analyzer. We can thus account
for the constant component of the background. An accurate solution
of the problem of combining the polarization given a successive
passage of light through a number of partially polarizing media
can be obtained by multiplying the matrices, transforming the
Stokes parameters for these media \citep{serk:Afanasiev_n,
shurk:Afanasiev_n}. Consider the case of finding the parameters of
linear polarization with a rotating phase plate $\lambda/2$ and a
polarizer, which passes only one projection (component) of the
electric vector. The components of the Stokes vector of radiation,
transmitted through such a system can be found from the relations:
\begin{equation}
\left|
\begin{array}{c}
I\arcmin\\
Q\arcmin\\
U\arcmin\\
V\arcmin\\
\end{array}
\right|
=
{1\over2}
M
\left|
\begin{array}{cccc}
1&0&0&0\\
0&{C_\lambda^2}-{S_\lambda^2}&2{C_\lambda}{S_\lambda}&0\\
0&2{C_\lambda}{S_\lambda}&{ S_\lambda^2}-{ C_\lambda^2}&0\\
0&0&0&-1\\
\end{array}
\right| \times \left|
\begin{array}{c}
I\\
Q\\
U\\
V\\
\end{array}
\right|, \label{equ1:Afanasiev_n}
\end{equation}
where
\begin{equation}
M=
\left|
\begin{array}{cccc}
1&C&S&0\\
C&C^2&CS&0\\
S&CS&S^2&0\\
0&0&0&0\\
\end{array}
\right|
\end{equation}
is the matrix of  transformation of the polarizer. Here
$$
\begin{array}{rr}
C=\cos(2\varphi), & S=\sin(2\varphi),\\
\\
{C_\lambda}=\cos(2\theta), & {S_\lambda}=\sin(2\theta),\\
\end{array}
$$
where  $\varphi$ is the direction of the polarizer  transmission
axis in the arbitrary coordinate system, and $\theta$ is the axial
direction of 
of the phase plate fast axis. In general, these directions do not
coincide.

Multiplying the matrices we obtain the Stokes parameters at the
output of the system:
\begin{equation}
\left|
\begin{array}{c}
I\arcmin\\
Q\arcmin\\
U\arcmin\\
V\arcmin\\
\end{array}
\right|
=
{1\over2}
\left|
\begin{array}{cccc}
I+C{Q_\lambda}  +S{U_\lambda}\\
IC+C^2{Q_\lambda}  +CS{U_\lambda}\\
IS+CS{Q_\lambda}  +S^2{U_\lambda}\\
0\\
\end{array}
\right|, \label{equ2:Afanasiev_n}
\end{equation}
where
$$
\begin{array}{cc}
{Q_\lambda}=Q({C_\lambda}^2-{S_\lambda}^2)+2U{C_\lambda}{S_\lambda},\\
\\
{U_\lambda}=Q({C_\lambda}^2-{S_\lambda}^2)+2U{C_\lambda}{S_\lambda}.\\
\end{array}
$$

The detector registers the intensity
\begin{equation}
\begin{array}{rl}
I\arcmin=\displaystyle\frac{1}{2}\{I+C[Q({C_\lambda}^2-{S_\lambda}^2)+2U{C_\lambda}{S_\lambda}] & \\
\\
+S[2Q{C_\lambda}{S_\lambda}-U({C_\lambda}^2-{S_\lambda}^2)]\}&\!\!.\\
\end{array}
\label{equ3:Afanasiev_n}
\end{equation}
To determine the linear polarization (the degree of polarization
$P$ and angle of the polarization plane $\phi$), we need two
Stokes $Q$ and $U$  parameters, normalized to the total intensity,
and related with $P$ and $\phi$ by known relations:
\begin{equation}
P=\sqrt{Q^2+U^2},\quad\phi=\frac{1}{2}\arctan{\frac{U}{Q}}~.\\
\end{equation}
Ideally, three measurements are required to determine the
parameters of linear polarization. In reality though, to reduce
the errors introduced by the imperfections of the measurement
equipment and the atmospheric noise, eight measurements are made:
for the $\theta$ angles equal to $0\degr$, $45\degr$, $22\fdg5$
and $67\fdg5$. In addition, for each of the these phase plate
orientation angles, the measurements of two angles of the
polarizer $\varphi$ and \mbox{$\varphi +\pi/2$} are made.

As a result of measurements in two angles of the phase plate
($0\degr$ and $45\degr$) given the polarization filter angles
$\varphi$ and \mbox{$\varphi+\pi/2$}, we register four
intensities:
\begin{equation}
\begin{array}{rcl}
I'_{0,\varphi}&=&\displaystyle{ {1\over2}}(I+CQ-SU),\vspace{-5mm}\\
\\
\\
I'_{0,\varphi+\pi/2}&=&\displaystyle{{1\over2}}(I-CQ+SU),\vspace{-5mm}\\
\\
\\
I'_{45,\varphi}&=&\displaystyle{{1\over2}}(I-CQ+SU),\vspace{-5mm}\\
\\
\\
I'_{45,\varphi+\pi/2}&=&\displaystyle{{1\over2}}(I+CQ-SU).\\
\end{array}
\end{equation}
Let us introduce the dimensionless quantities
\begin{equation}
\begin{array}{rclcl}
F_0&=&\frac{\displaystyle{I'_{0,\varphi}-I'_{0,\varphi+\pi/2}}}
{\displaystyle{I'_{0,\varphi}+I'_{0,\varphi+\pi/2}}}&=&
\frac{\displaystyle{Q\cos{2\varphi}-U\sin{2\varphi}}}{\displaystyle{I}},\vspace{-5mm}\\
\\
\\
F_{45}&=&\frac{\displaystyle{I'_{45,\varphi}-I'_{45,\varphi+\pi/2}}}
{\displaystyle{I'_{45,\varphi}+I'_{45,\varphi+\pi/2}}}&=&
\frac{\displaystyle{U\sin{2\varphi}-Q\cos{2\varphi}}}{\displaystyle{I}}.\\\end{array}
\end{equation}
If the phase plate shifts the phases of the ordinary and
extraordinary waves exactly by $\lambda/2$ in the entire
wavelength range and the angles of orientation of the polarization
filter and the phase plate are absolutely accurate, then
\mbox{$F_0\equiv-F_{45}$}. Really small errors in the intensity
measurements for a given set of angles introduce the errors in
calculating $F_0$ and $F_{45}$ with the opposite sign. Hence a
combination
\begin{equation}
F_{0}-F_{45}=2\frac{\displaystyle{Q\cos{2\varphi}-U\sin{2\varphi}}}{\displaystyle{I}}
\end{equation}
allows to reduce the effects of nonideality of the system's
optical elements and thus significantly improve the accuracy of
measurements.

All of the above applies to the measurements in the angles of
$22\fdg5$ and $67\fdg5$ degrees, and by analogy with the
derivation of (5) and (6) we can write
\begin{equation}
F_{22.5}-F_{67.5}=2\frac{\displaystyle{Q\sin{2\varphi}+U\cos{2\varphi}}}{\displaystyle{I}}.
\end{equation}
In fact, adding the measurements in the angles $45\degr$ and
$67\fdg5$, we use the modulation ideology to reduce the effect of
errors of the optical path on the results of measurements of
linear polarization. If a Wollaston prism is used as the analyzer,
we obtain simultaneous measurements in mutually perpendicular
directions $\varphi$ and $\varphi+\pi/2$, which is equivalent to
the application of  two polarizers.

For measuring the circular polarization, the phase plate
$\lambda/4$ is inserted in the beam, shifting the phase by
\mbox{$\pm90\degr$}. In this case, the Stokes parameters at the
output of the system can be written as follows:
\begin{equation}
\left|
\begin{array}{c}
I\arcmin\\
Q\arcmin\\
U\arcmin\\
V\arcmin\\
\end{array}
\right|
=
{1\over2}
M
\left|
\begin{array}{cccc}
1&0&0&0\\
0&{C_\lambda^2}&{C_\lambda}{S_\lambda}&-{S_\lambda}\\
0&{C_\lambda}{S_\lambda}&{ S_\lambda^2}&{C_\lambda}\\
0&{S_\lambda}&-{C_\lambda}&-1\\
\end{array}
\right| \times \left|
\begin{array}{c}
I\\
Q\\
U\\
V\\
\end{array}
\right|. \label{equ1:Afanasiev_n}
\end{equation}
The measurements in two angles of the phase plate $\varphi=0\degr$
and $90\degr$  for two orientations of the polarizer $0\degr$ and
$90\degr$ yield four intensity measurements:
\begin{equation}
\begin{array}{rcl}
I'_{0,\varphi}&=&\displaystyle{{1\over2}}(I+CQ+SV),\vspace{-5mm}\\
\\
\\
I'_{0,\varphi+\pi/2}&=&\displaystyle{{1\over2}}(I-CQ+SV),\vspace{-5mm}\\
\\
\\
I'_{90,\varphi}&=&\displaystyle{{1\over2}}(I+CQ-SV),\vspace{-5mm}\\
\\
\\
I'_{90,\varphi+\pi/2}&=&\displaystyle{{1\over2}}(I-CQ+SV).\\
\end{array}
\end{equation}
Whence it follows
\begin{equation}
F_{0}-F_{90}=2\frac{\displaystyle{V\sin{2\varphi}}}{\displaystyle{I}}.
\end{equation}

From relations (7), (8) and (11) it is clear that if we set in a
certain way the fast axes of the phase plates relative to the
direction of the unit vectors of the ordinary {\bf o} and
extraordinary {\bf e} rays in the case when a Wollaston prism is
used as an analyzer, we can get simple relations between the
combinations of  dimensionless quantities $F_j$ and the Stokes
parameters. It is obvious that the direction of the fast axis of
the $\lambda/2$-plate should coincide with the unit vector {\bf
o}, and the \mbox{$\lambda/4$-plate} has to be oriented at an
angle of $45\degr$ to the unit vector {\bf o}. This fact will be
used in what follows when processing the polarimetric data. In
real observations, when the observed polarization of the object is
distorted by the instrumental polarization, the Earth's atmosphere
and the interstellar medium, the Stokes vector we measure can be
represented as a result of vector summation of the real Stokes
vector with the vectors of instrumental polarization,
depolarization in the Earth's atmosphere and the interstellar
medium~\citep{trip:Afanasiev_n}. The calculations show
\citep{shakh1:Afanasiev_n,shakh2:Afanasiev_n} that the measured
Stokes parameters can be represented as a linear combination of
the Stokes parameters of the above mentioned vectors with the
accuracy of 0.025\% when the degree of polarization of the object
is not exceeding 5\%. For large polarizations (of about 10\%), the
error can reach 0.2\%. The exact formulas should be used only in
the analysis of high-precision observations of objects with a
large (over 10\%) polarization. The instrumental polarization
introduced by the telescope and the measuring equipment may be
taken into account after the ad hoc observations of unpolarized
standard stars. The technique of such corrections is described in
detail  in the surveys by Serkowski and
McCarthy~\citep{serk:Afanasiev_n,mac:Afanasiev_n}.

\section{TECHNIQUE OF POLARIMETRIC DATA REDUCTION}

The spectropolarimetric data reduction technique is different for
the observations of objects in three operating modes of the
SCORPIO spectrograph:
\begin{list}{}{
\setlength\leftmargin{2mm} \setlength\topsep{1mm}
\setlength\parsep{-0.5mm} \setlength\itemsep{2mm} }

\item 1) the observations with a single Wollaston prism (WOLL-1)
and the phase plates, rotating at fixed angles with the
$\lambda/2$ and $\lambda/4$ phase shifts;

\item  2) the observations with a double Wollaston prism (WOLL-2);

\item  3) the observations with a rotating dichroic analyzer of
linear polarization (POLAROID).
\end{list}

The use of a particular mode depends on the task and astroclimatic
conditions. The  characteristic property of spectropolarimetric
observations is that the Mueller matrix, describing the
transformations of the Stokes vector in the registration system,
can always be diagonalized in sufficiently narrow spectral
intervals, i.e., the measured values of the Stokes parameters are
a linear combination of the intrinsic parameters. This feature
distinguishes the spectropolarimetric observations from the
broadband photometry. In the latter case, the Mueller matrix has a
triangular shape inside the photometric band, and hence the system
has the instrumental polarization which is difficult to account
for.

\subsection{Single Wollaston Prism}

\begin{figure}

\includegraphics[width=\columnwidth]{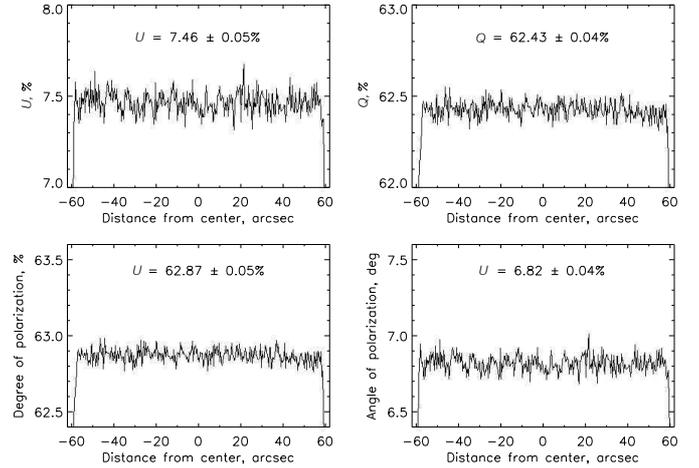}
\caption{The variations of the Stokes $Q$ and $U$ parameters, the
degree of polarization and the angle of the polarization plane
based on the observations of the twilight sky.}
\label{instr:Afanasiev_n}
\end{figure}

In this mode, the Wollaston prism, separating the beam into the
ordinary {\bf o} and extraordinary {\bf e} rays by $5\degr$
(2\arcmin\ on the celestial sphere), is inserted in the parallel
beam. The {\bf o} and {\bf e} rays are the projections of the
input polarization vector {\bf E} in two mutually perpendicular
directions. The prism is mounted in the spectrograph in such a
manner that the direction of the unit vector {\bf o} coincides
with an accuracy of about $0\fdg05$ with the direction of the
spectrograph slit. The achromatic phase plates are mounted near
the focal plane of the BTA in the divergent beam. This design
concept of the phase plate positioning, realized in the SCORPIO-2,
in contrast to the FORS spectrograph of the \mbox{8-m} VLT
telescope~\citep{apen:Afanasiev_n,foca:Afanasiev_n} and the FOCAS
spectrograph of the \mbox{10-m} SUBARU  telescope
~\citep{kawa:Afanasiev_n} is advantageous in that it does not
introduce any significant instrumental polarization along the slit
height (see Fig.~\ref{instr:Afanasiev_n}). This is due to the fact
that the  phase plates in the FORS and the FOCAS are mounted in
the parallel beam, and they have, like any device with a
multi-layer coating, the dependence of the phase shift on the
angle of incidence. Therefore, since both instruments are the
telescopic systems, the phase shift depends on the object's
position along the slit height. As noted
in~\citep{fossa:Afanasiev_n}, the hence introduced instrumental
polarization can reach up to 1.5--1.6\% at a distance of
2\arcmin{} from the slit center.

In the observations with a single Wollaston prism, we  register
two spectra with the intensity of $I_{o}(\lambda)$   and
$I_{e}(\lambda)$ (ordinary and extraordinary, respectively). In
general, the orientation of the principal axes of the prism
$\varphi$ and the directions of the fast axis vector  $\theta$ of
the phase plates are arbitrary. Then, as follows from Section 3,
the  values of  the measured, normalized to the integral intensity
$I$ Stokes   parameters $U$ and $Q$
would be found from the relations:
\begin{equation}
\begin{array}{rcl}
Q'&=&\left[R(\lambda)_{\theta=0\degr}-R(\lambda)_{\theta=45\degr}\right]/2,\\
\\
U'&=&\left[R(\lambda)_{\theta=22.5\degr}-R(\lambda)_{\theta=67.5\degr}\right]/2,\\
\end{array}
\end{equation}
where
$R(\lambda)=\biggl(\displaystyle{\frac{I_o(\lambda)-I_e(\lambda)}{I_o(\lambda)+I_e(\lambda)}}\biggl)$,
$\theta$ is the angle of rotation of the $\lambda/2$ phase plate,
and the intrinsic values of the second and third Stokes parameters
are found from the rotation transformation:
\begin{equation}
\begin{array}{rcl}
Q/I&=& Q'\cos{2\varphi}- U' \sin{2\varphi},\\
\\
U/I&=& Q'\sin{2\varphi}+ U' \cos{2\varphi}.\\
\end{array}
\end{equation}
The degree of linear polarization $P$ is determined from relation
(5), and the observed position angle of the polarization plane
${\rm PA}$ is calculated as follows:
 \begin{equation}
{\rm PA}={\rm PA}_{\rm
slit}-\displaystyle{\frac{1}{2}}\arctan{\displaystyle{\frac{U}{Q}}
}+{\rm PA}_0.
\end{equation}
Here  ${\rm PA}_{\rm slit}$ is the position angle of the major
axis of the Wollaston prism, which coincides with the direction of
the slit, and ${\rm PA}_{0}$ is the zero point of the dependence
of the slit position angle on the angle of the BTA primary focus
turntable and the parallactic angle of the object. Note that the
minus sign in the formula is due to the fact that the image at the
entrance of the spectrograph is a reflection of the image on the
celestial sphere.

For the measurements of circular polarization, the $\lambda/4$
plate is inserted in the beam, which is oriented in such a way
that the direction of the fast axis makes an angle of $45\degr$
with the direction of the main axis of the Wollaston prism. The
plate turns by two fixed angles, $0\degr$ and $90\degr$. In this
case the value of the circular polarization (the Stokes $V$
parameter) is found from the relation:
\begin{equation}
V/I=\left[R(\lambda)_{\theta=0\degr}-R(\lambda)_{\theta=90\degr}\right]/2.
\end{equation}
It is commonly known that this measurement technique allows to
eliminate the phase shift errors of the phase plates, but requires
a stable atmosphere. In reality though several (more than three)
cycles of measurements are carried out during the observations,
when we successively record the spectra for pairs of angles
($0\degr$, $45\degr$) and ($22\fdg5$, $67\fdg5$) for the
$\lambda/2$ plate and ($0\degr$, $90\degr$) for the $\lambda/4$
plate. The duration of the exposure should be at least 3--4 times
longer than the CCD frame readout time.

\subsection{Double Wollaston Prism}

A double Wollaston prism, the concept of which was first proposed
and realized by Geyer et al.~\citep{gey:Afanasiev_n} and developed
by Oliva~\citep{oliva:Afanasiev_n} features two Wollaston prisms
inserted at the exit pupil of the spectrograph. One of the prisms,
with the directions of polarization axes of $0\degr$ and
$90\degr$, is illuminated by one half of the exit pupil, and the
other prism, with the directions of axes of $45\degr$ and
$135\degr$ is illuminated by the other half of the pupil. Two
achromatic wedges were used to separate the images from each
prism. Thus, using this registration method, we obtain four
spectra with the intensities of $I(\lambda)_{0}$,
$I(\lambda)_{90}$, $I(\lambda)_{45}$,  and $I(\lambda)_{135}$.

Then,
the values of the second and third normalized Stokes $Q$ and $U$
parameters are determined from the relations:

\begin{equation}
\begin{array}{rcl}
Q'&=&\displaystyle{\frac{I(\lambda)_{0}-I(\lambda)_{90}}{I(\lambda)_{0}+I(\lambda)_{90}}},\vspace{-5mm}\\
  \\
  \\
U'&=&\displaystyle{\frac{I(\lambda)_{45}-I(\lambda)_{135}}{I(\lambda)_{45}+I(\lambda)_{135}}}.\\
\end{array}
\end{equation}
For the measurement of circular polarization the $\lambda/4$ phase
plate is inserted in the beam and then
\begin{equation}
\begin{array}{rl}
V/I=0.5\biggl(\displaystyle{\frac{I(\lambda)_{0}-I(\lambda)_{90}}{I(\lambda)_{0}+I(\lambda)_{90}}}\biggr)&\vspace{-5mm}\\
\\
\\
-0.5\biggl(\displaystyle{\frac{I(\lambda)_{45}-I(\lambda)_{135}}{I(\lambda)_{45}+I(\lambda)_{135}}}\biggr)&\!\!.\\
\end{array}
\end{equation}

The values  of the Stokes $U$ and $Q$  parameters, and the
position angle of the polarization plane ${\rm PA}$ are calculated
here using the (14) and (15) relations, and the degree of
polarization $P$ is found from formula (5).

This measurement method, as opposed to the previous one, provides
for a  simultaneous measurement of the  Stokes $Q$ and $U$
parameters, which is important in the conditions of unstable
atmosphere. However, its disadvantage is the need to calibrate the
sensitivity of four independent polarization channels by the zero
polarization standard. Although the calibration dependences
$I(\lambda)_{0}/I(\lambda)_{90}$ and
$I(\lambda)_{45}/I(\lambda)_{135}$  are constant for the
instrument, nevertheless, they have to be refined on each night of
observations.

\subsection{Dichroic Polarization Analyzer}

The dichroic polarization analyzer (also called the polarization
filter or the polaroid), mounted in the spectrograph is intended
for the measurements of linear polarization in starlike and
extended objects in the field of view with the diameter of
6\arcmin. It is also planned to be used for two-dimensional
spectropolarimetry after the commissioning of the integral field
spectroscopy mode at the SCORPIO-2 spectrograph.

The analyzer can be set in three fixed positions by the angle,
$0\degr$ and $\pm 60\degr$. Then, with intensities measured in
three angles of the polaroid, $I(x,y)_{0\degr}$,
$I(x,y)_{-60\degr}$ and $I(x,y)_{+60\degr}$ we can calculate, up
to the rotation transform, the measured Stokes $Q'$ and $U'$
parameters in each point of the image with the coordinates
$(x,y)$:
\begin{equation}
\begin{array}{rcl}
Q'(x,y)&=&\displaystyle{\frac{2I(x,y)_{0}-I(x,y)_{-60}-
I(x,y)_{+60}}{I(x,y)_{0}+ I(x,y)_{-60}+ I(x,y)_{+60}}},\vspace{-5mm}\\
  \\
  \\
U'(x,y)&=&\displaystyle{\frac{1}{\sqrt{3}}\frac{I(x,y)_{+60}-I(x,y)_{-60}}
{I(x,y)_{0}+ I(x,y)_{-60}+ I(x,y)_{+60}}}.\\
\end{array}
\end{equation}
The values of the Stokes $Q$ and $U$  parameters, the degree of
linear polarization $P$ and the position angle of the polarization
plane ${\rm PA}$ are calculated here from relations (14), (5) and
(15), respectively. This measurement technique was proposed by
Fesenkov, but out of three observed intensity values he only
computed the degree of polarization and the rotation angle of the
polarization plane ~\citep{mart:Afanasiev_n}. Just like in the
case of the single Wollaston, this measurement technique depends
on the atmospheric conditions.

 \subsection{Generation of the Initial Data Cube}

A common problem encountered in the reduction of polarimetric
observations is that when we compare the differential values in
each image point, any, even very small differences between the
images obtained at different angles of the phase plates or the
polaroid, which are introduced at various stages of the reduction,
impair the accuracy of polarization measurements. The same applies
to the removal of the traces of cosmic ray particles by different
smoothing techniques, since these  algorithms typically shift the
statistic estimates, and therefore contribute some depolarization
effects to the final result.

A large amount of different images is generated, obtained at
various angles of phase plates, possessing different features
({\tt obj}, {\tt flat}, {\tt bias}, {\tt neon}, {\tt dark}, {\tt
etalon}) and types of observed objects (the studied object, the
zero polarization standard, the non-zero polarization standard).
This calls for a uniformity of the input data fed to the reduction
software with the minimal intervention of the astronomer. To this
end, the reduction software first analyzes the
\mbox{FITS-headings} of the input array of images and generates
the following data:
\begin{list}{}{
\setlength\leftmargin{1mm} \setlength\topsep{1mm}
\setlength\parsep{-0.5mm} \setlength\itemsep{2mm} }

\item 1) $\tt obj(x, y, pol, exp)$ is the 4-dimensional cube of
initial data, where $\tt x,y$ are coordinates of the point on the
image, $\tt pol$ is the identifier of the phase plate and its
angle:
$$
\begin{array}{ll}
{\tt pol}=0\rightarrow\lambda/2\,(0\degr), & {\tt pol}=1\rightarrow\lambda/2\,(45\degr),\\
{\tt pol}=2\rightarrow\lambda/2\,(22\fdg5), & {\tt pol}=3\rightarrow\lambda/2\,(67\fdg5),\\
{\tt pol}=4\rightarrow\lambda/4\,(0\degr), & {\tt pol}=5\rightarrow\lambda/4\,(90\degr),\\
\multicolumn{2}{c}{{\tt pol}=6\rightarrow {\rm polaroid}\,(0\degr),\phantom{+6}} \\
\multicolumn{2}{c}{{\tt pol}=7\rightarrow {\rm polaroid}\,(+60\degr),} \\
\multicolumn{2}{c}{{\tt pol}=8\rightarrow {\rm polaroid}\,(-60\degr);} \\
\end{array}
$$
and ${\rm exp}$ is an identifier that determines the type of the
object and the number of the exposure;

\item 2) $\tt bias(x,y)$ is a robust estimate of the electrical
zero  of the registration system;

\item 3) $\tt neon(x,y,pol)$ is the image of the line spectrum for
the wavelength scale calibration;

\item 4) $\tt flat(x,y,pol)$ is the image of the spectra of a
continuous-spectrum lamp used for the transmission calibration.
\end{list}

The subsequent data processing steps involve the construction of a
model of scale distortions, reduction of data for these
distortions and the sensitivity heterogeneity, the sky background
subtraction, the wavelength scale calibration, extraction of the
spectra from the images (or the integral fluxes from the objects,
given the field photometry), and finally, the calculation of the
Stokes parameters. The removal of the traces of cosmic ray
particles is done at the final stage of reduction via the robust
parameter estimates.

\subsection{Correction of Scale Distortions}

The images obtained in the 2D-spectropolarimetry mode are strongly
distorted for various reasons:
\begin{list}{}{
\setlength\leftmargin{2mm} \setlength\topsep{1mm}
\setlength\parsep{-0.5mm} \setlength\itemsep{2mm} }

\item 1) the spectral line curvature along the slit height, caused
by the variation of the meridional increase of dispersion elements
used in the spectrograph, namely, the Volume Phase Holographic
Gratings (VPHG) combined with prisms;

\item  2)  the chromatism of the Wollaston prism, leading to the
wavelength-dependent value of the divergence of ordinary and
extraordinary rays, which increases the curvature of spectral
trajectories, as compared with the conventional spectroscopy;

\item  3)  the distortion of the spectrograph camera optics and
the higher order spherical chromatic aberrations also leading to
the geometric distortions.
\end{list}

The value of geometric distortions in our case can reach up to
10--15\% in the observations covering a wide spectral range. Given
that in the case of accurate measurements the errors of the Stokes
parameters should not exceed 0.1\%, the final coordinate-wise
nonlinearity of data in the spectra
cannot exceed 0.1~px, what
corresponds to  about 0.005\% for the CCD we use
(\mbox{$2068\times4632$}~px). In order to obtain such a high
accuracy, the scale distortions have to be carefully calibrated,
and exact algorithms of coordinate conversion have to be applied.

As a model of geometric transformations, we use the geometric
polynomial $N$-order transformation, where the resulting set of
intensity values is determined by the  following relations:
\begin{equation}
I[x,y]=I'[x',y']= I'[a(x,y),b(x,y)]~,
\end{equation}
where  $I[x,y]$ is the intensity in the point with the coordinates
$(x,y)$ in the undistorted image, which corresponds to the
intensity   $I'[x',y']$, measured in the point of the distorted
image with the coordinates $(x',y')$.

The functions $a(x,y)$ and $b(x,y)$ are the polynomials of the
$N$th degree, whose coefficients  $K_{x_{i,j}}$ and $K_{y_{i,j}}$
determine the spatial transformation:
\begin{equation}
x'=a(x,y)=\sum_{i=0}^N \sum_{j=0}^N K_{x_{i,j}}x^j y^i
\end{equation}
\begin{equation}
y'=b(x,y)=\sum_{i=0}^N \sum_{j=0}^N K_{y_{i,j}}x^j y^i.
\end{equation}

The accuracy of the model of geometric distortions  depends on the
accuracy of determining the  $K_{x_{i,j}}$ and $K_{y_{i,j}}$
coefficients. From this point on, we consider that the degree of
the polynomial, responsible for our distortions, is not higher
than three, and the resulting coordinates of nodes are accurate,
so that the spline interpolation could be used in the geometric
transformation. It should also be noted that the accuracy of the
geometrical model determines the quality of the night sky spectrum
subtraction, which is important for the polarization observations
of faint objects.

The model of distortions is constructed as follows.

\begin{figure}
\includegraphics[width=\columnwidth]{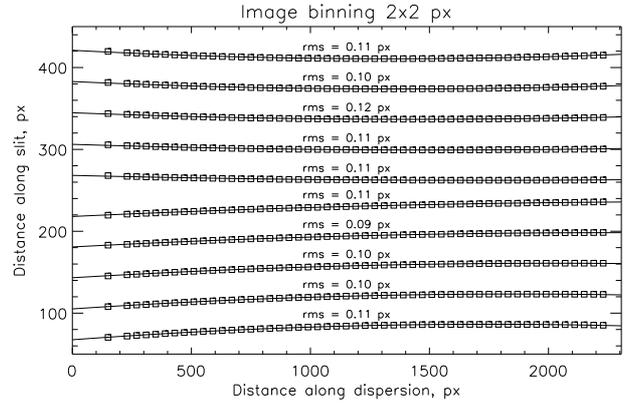}
\caption{The trajectories of the
etalon spectra for the single Wollaston prism.}
\label{eta:Afanasiev_n}
\end{figure}

First, based on the analysis of the etalon  spectra (which is an
image of the individual spectra of the continuous-spectrum lamp,
given that a dot mask is inserted in front of the slit) the
curvature of the spectra along dispersion is determined, which we
shall further call the trajectories of the spectra.

1)  First, based on the analysis of the etalon spectra
(which is an image of individual spectra of a continuous-spectrum
lamp, given that a dot mask is inserted in front of the slit) the
curvature of the spectra along dispersion is determined, which we
shall further call the trajectories of the spectra. For the single
and double Wollaston prisms, a 5- and a 3-dot mask is inserted in
front of the slit, respectively. Figure~\ref{eta:Afanasiev_n}
shows an example of construction of the trajectories for the image
of the standard obtained with a single Wollaston prism. The
positions of nodes are determined from the analysis of spectral
sections across dispersion. The positions of the obtained points
are approximated by the polynomial of the third degree, which
gives the true shape of the trajectories.

\begin{figure*}
\begin{minipage}[t]{1.0\linewidth}

\includegraphics[scale=1]{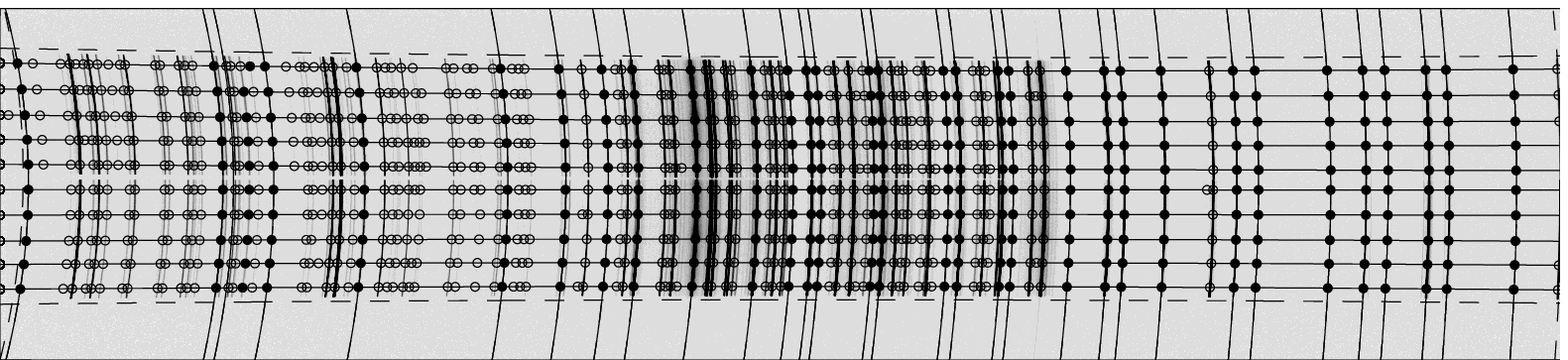}
\caption{The approximation of the curvature of the comparison
spectrum lines by the third-degree polynomials for the single
Wollaston prism.} \label{lines:Afanasiev_n}
\vspace{5mm}
\end{minipage}
\begin{minipage}[t]{1.0\linewidth}

\includegraphics[scale=1]{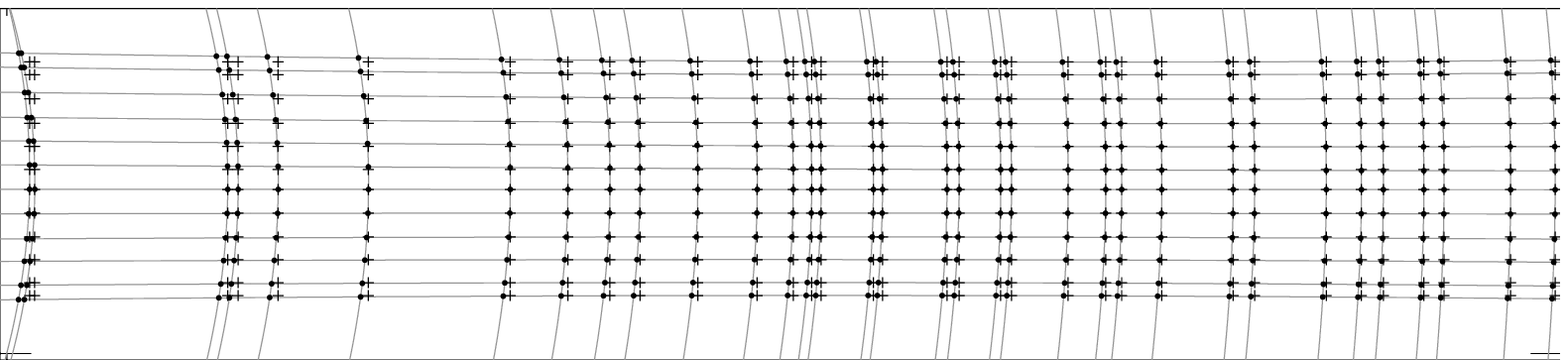}
\caption{The nodes of the grid of the geometric distortion model.
The crosses and points mark the undistorted and distorted images,
respectively. } \label{grid:Afanasiev_n}
\vspace{5mm}
\end{minipage}
\begin{minipage}[t]{1.0\linewidth}
\includegraphics[scale=0.8]{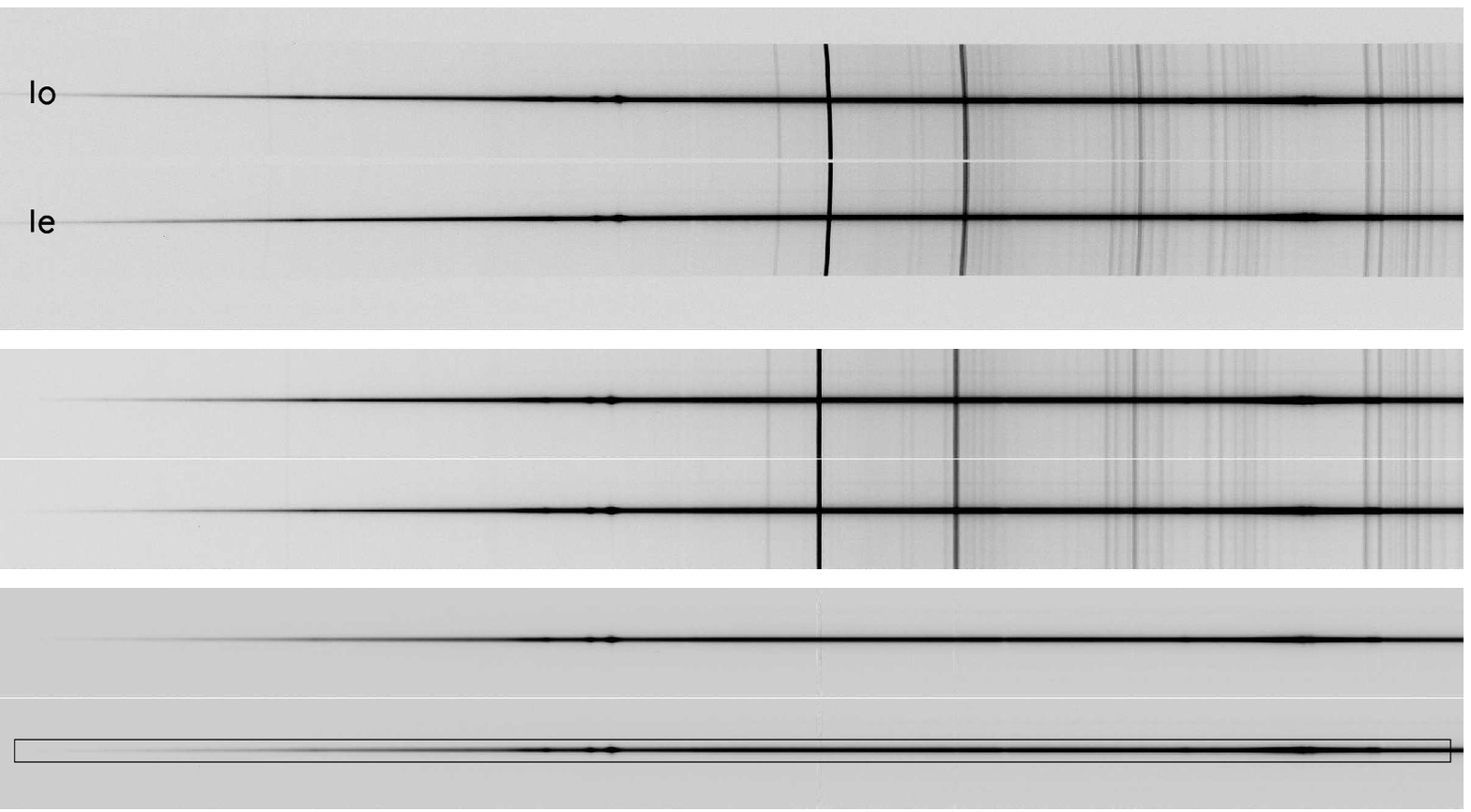}
\caption{Image processing with a single Wollaston prism: the top
plot represents the original spectrum, the middle plot---the
spectrum, corrected for the geometric distortions, and the bottom
plot presents the spectrum after the background sky subtraction.}
\label{2D:Afanasiev_n}
\end{minipage}
\end{figure*}

2)  Then, the obtained trajectories are superimposed over the
two-dimensional image of the comparison spectrum to determine the
positions of spectral lines at the nodes of their intersections
with the trajectories. Out of the entire array of the obtained
intersections we determine the points, coinciding by wavelength
for each trajectory.
Based on them, a model of the curvature along the slit height is
constructed for each line using the third-degree polynomial. An
example of identification of spectral lines and the approximation
of their curvature by the third-degree polynomials is shown in
Fig.~\ref{lines:Afanasiev_n}. At that, an individual polynomial is
constructed for each identified line.

3) At last, at the final stage the obtained coefficients of the
polynomials, defining the curvature of the lines and trajectories
are approximated by wavelength by the second-degree polynomial,
and based on the results of approximation of these coefficients
the smoothed curved lines and trajectories of the spectra are
constructed.
Next, the  points of intersection $(x',y')$ of line trajectories
are computed.  Here, the $x$-coordinates of lines and the
$y$-coordinates of the trajectory in the center of the image are
regarded as the true (undistorted) coordinates $(x,y)$. The exact
coordinates of the nodes obtained by these means are used to
determine the two-dimensional polynomial coefficients, found from
the expressions (21) and (22). At the edges of the image of the
spectrum the nodes are extrapolated according to the approximation
of the curvature of lines and trajectories. The result of
constructing the nodes of the geometrical model grid in the image
is demonstrated in Fig.~\ref{grid:Afanasiev_n}.

Next, all the original images ({\tt obj}, {\tt flat}, {\tt neon})
are corrected for the geometric distortions using the IDL
environment WARP\_TRI procedure, the corrections for the
sensitivity heterogeneity are done (the flat-field procedure) and
the sky background is subtracted. Our numerous tests have shown
that the actual accuracy of the correction for geometric
distortions is at least 0.2~px, what corresponds to $3\mu$m in the
plane of the CCD. The examples of images at various stages of
reduction of data, obtained with the single Wollaston prism are
shown in Fig.~\ref{2D:Afanasiev_n}.

The wavelength scale calibration is performed is a standard way:
automatic line identification, two-dimensional approximation of
the dispersion curve by the third-order polynomial, quadratic
smoothing of the polynomial coefficients along the slit height,
and image linearization.

\subsection{Extraction of Spectra and Calculation of the Stokes Parameters}

\begin{figure*}

\includegraphics[scale=1]{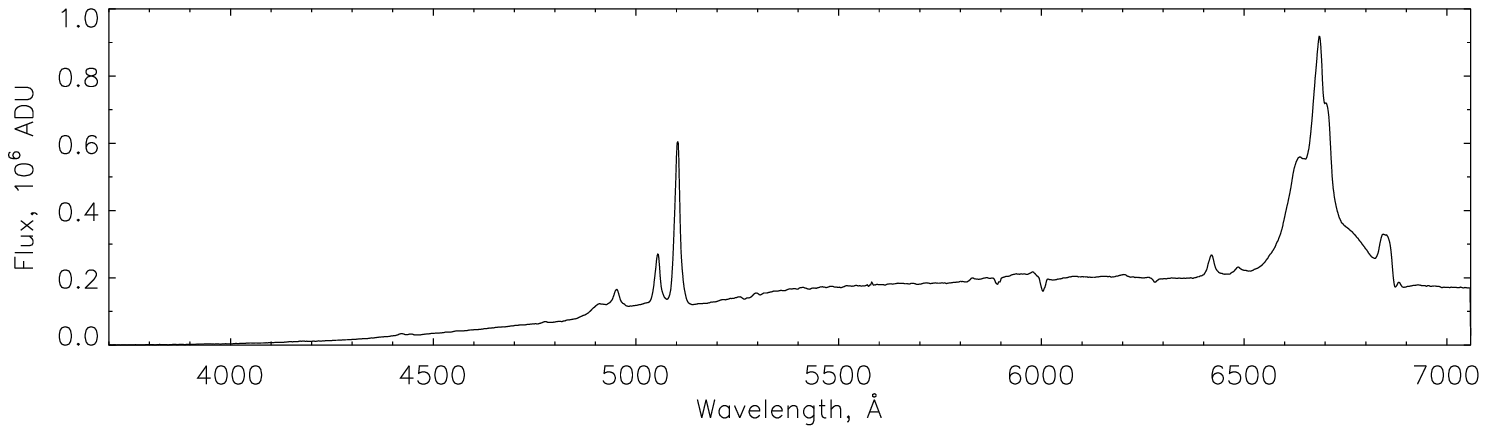}
\caption{The spectrum of the nucleus of the Markarian~6 galaxy,
extracted from the image in Fig.~\ref{2D:Afanasiev_n}. }
\label{sp:Afanasiev_n}
\end{figure*}

The accuracy of calculation of the Stokes parameters, besides the
exactness of geometric transformations, depends on how the
dimensionless quantities $R(\lambda)$, participating in the
relations (13), (16), (17) and (18) are determined. In case of the
observations of extended objects, the above equations involve the
intensity values of the ordinary and extraordinary rays $I_{\rm
o}(x,y)$ and $I_{\rm e}(x,y)$,  which are used to calculate the
Stokes parameters at each point in the image. The mean Stokes
parameter value for each position along the slit (up to the image
size) is calculated by the robust methods using the stellar image
profile at the entrance of the spectropolarimeter as a scaling
function. However, in the case of starlike objects we have to use
the methods similar to the aperture photometry
techniques~\citep{stet:Afanasiev_n}, which allow to extract the
spectrum of the object from the image in an optimal way (in terms
of the maximum signal-to-noise ratio), and also do not introduce
any dummy instrumental polarization. Moreover, the impulse
interferences in the image, mainly brought in by the cosmic ray
particles, are efficiently suppressed. An example of the spectrum
extraction is shown in Fig.~\ref{sp:Afanasiev_n}.

For the optimal extraction of the spectra, the approximation of
the stellar profile along the slit (by the $y$-coordinate)
according to the Moffat model \citep{moff:Afanasiev_n} is
generally applied:

\begin{equation}
{\rm PSF}(y)=a_0+\displaystyle{\frac{a_1}{(u+1)^{a_6}}},
\end{equation}
where
$u=\displaystyle{\biggl(\frac{y-a_4}{a_2}\biggr)^2+\biggl(\frac{y-a_4}{a_2}\biggr)^2}$.
In more complex cases, the profile is approximated by a
combination of Gaussian functions.

\subsection{Atmospheric Condition Monitoring}

\begin{figure*}
\includegraphics[scale=0.7, bb=0 0 612 530,clip]{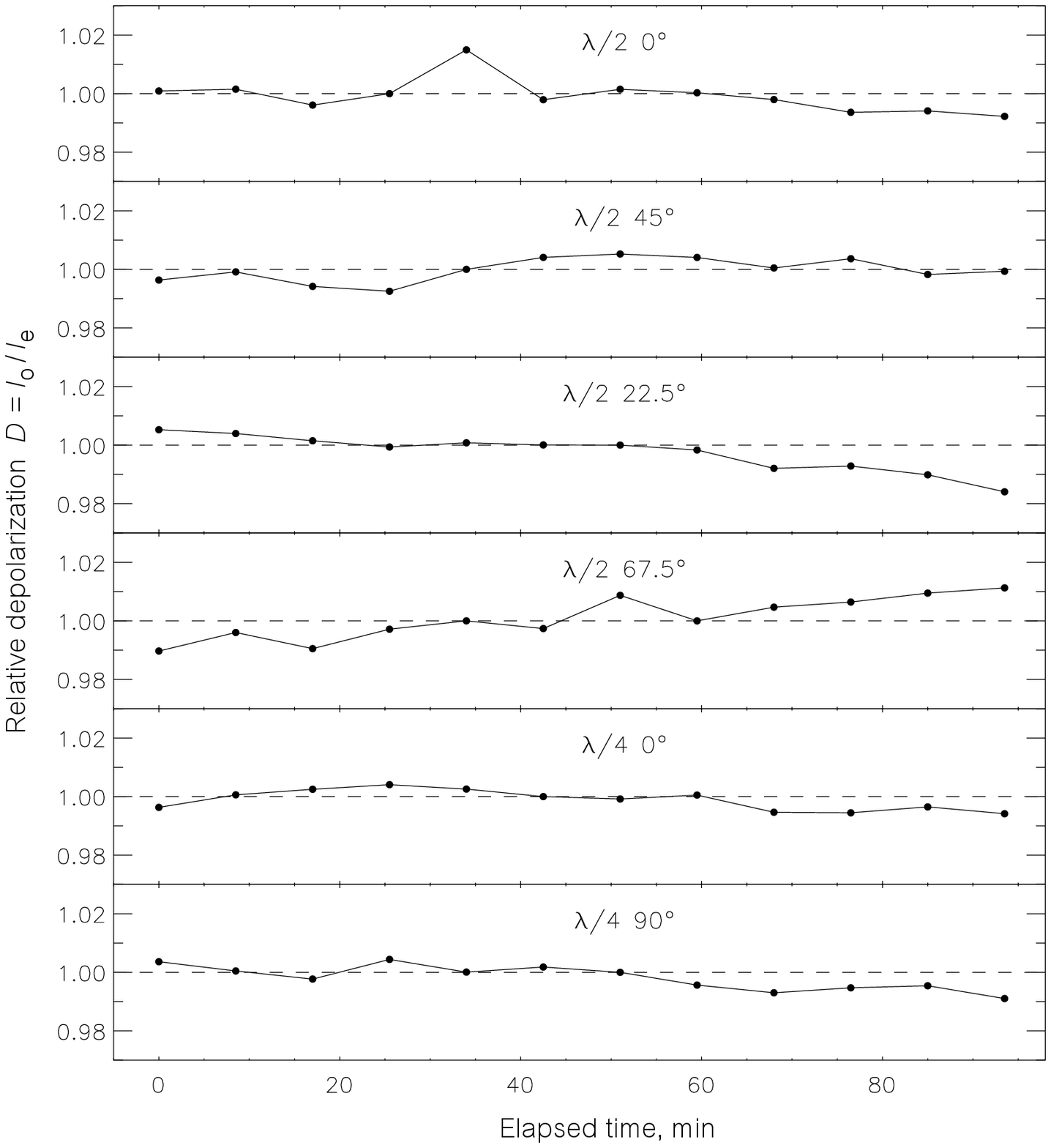}
\caption{The variation of  the depolarization coefficients
$D=I_o/I_e$ normalized for the median value for the successive
60-s exposures with the step of 100~s in the observations with a
single Wollaston prism. } \vspace{5mm} \label{atm:Afanasiev_n}
\end{figure*}

One of the most serious factors limiting the accuracy of the
ground-based polarization observations is the effect of the
Earth's atmosphere. A simple transparency variation in the case of
the double-beam scheme of polarization measurements has little
effect on the accuracy of measurement of the normalized Stokes
parameters. However, the variation of the parameters of the
radiative transfer in the atmosphere leads to an adjustment of the
depolarization coefficient $D=I_{\rm o}/I_{\rm e}$ at different
exposures, what is illustrated in Fig.~\ref{atm:Afanasiev_n}. The
main reason of depolarization is the selective light scattering
from single microparticles that are smaller than the wavelength of
the observed radiation (the Rayleigh scattering with magnitude
proportional to $\lambda^{-4}$), and non-selective aerosol light
scattering~\citep{chen:Afanasiev_n}.  In the case of observations
from the BTA location, where the atmosphere is dominated by the
aerosol scattering~\citep{kart:Afanasiev_n}, we can expect that
the effects of depolarization will  depend little on wavelength.
Moreover, the characteristic times of the fast depolarization
variations in the atmosphere are comparable with the the exposure
times short enough to ``freeze'' the  atmospheric turbulence,
and, depending on the atmospheric conditions
during the observations, amount to about \mbox{10--30~ms}. To
suppress the effects of depolarization (which are called the
scintillation effects) the schemes with a fast modulation or
polarization channel switching are generally applied. In
spectropolarimetry this is only possible with the use of the KDP
crystals and only for bright objects, since such schemes have an
extremely low transmission.

\begin{figure*}
\begin{minipage}[t]{1.0\linewidth}

\includegraphics[scale=0.9, bb=0 0 463 260,clip]{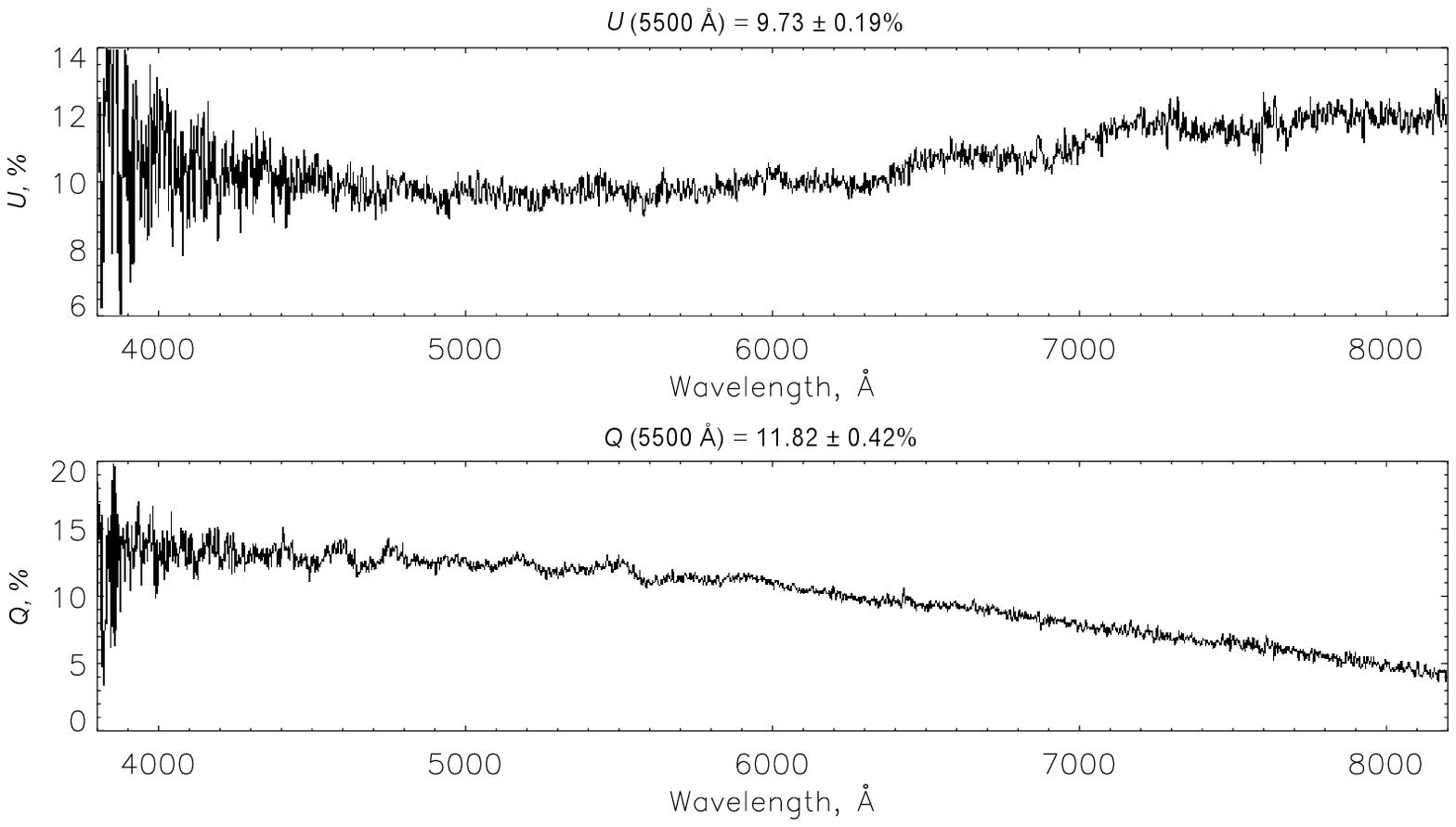}
\caption{The influence of depolarization effects on the wavelength
dependence of the Stokes parameters for the blazar S5\,0716+71. }
\label{uncorr:Afanasiev_n}
\end{minipage}
\begin{minipage}[t]{1.0\linewidth}

\includegraphics[scale=0.9, bb=0 0 463 260,clip]{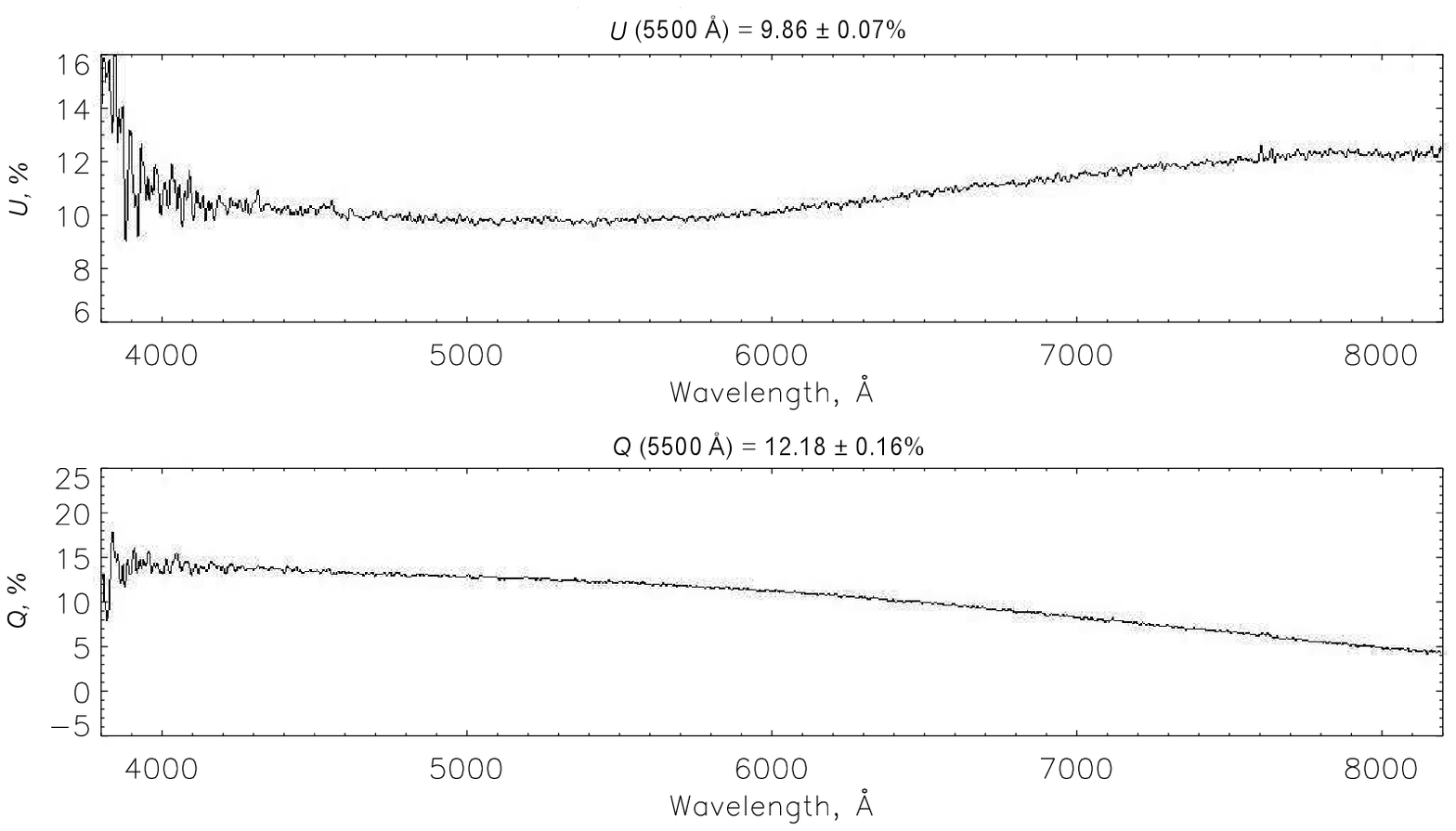}
\caption{The wavelength dependence of the Stokes parameters for
the blazar S5\,0716+71 after the correction for the depolarization
effects in the Earth's atmosphere.} \vspace{10mm}
\label{corr:Afanasiev_n}
\end{minipage}
\end{figure*}

\begin{table*}

\vspace{5mm} \caption{A comparison of the measured parameters of
linear polarization for WOLL-1} \label{globdata:Afanasiev_n}
\medskip
\begin{tabular}{l|r|r|r|r}
\hline
\multicolumn{1}{c|}{Star}&\multicolumn{1}{c|}{$P_{\rm obs}$,\%}&\multicolumn{1}{c|}{$P_{\rm tab}$,\%}&\multicolumn{1}{c|}{$\rm PA_{obs}$, deg}&\multicolumn{1}{c}{$\rm PA_{tab}$, deg}\\
\hline
BD\,+58$\degr$2272 & $5.52\pm0.07$ & $5.49\pm 0.02$  & $58.4\pm 0.7$   & $59.3\pm0.1$   \\
HD\,14433 & $3.65\pm0.07$ & $3.77\pm0.01$  & $113.7\pm0.5$     & $112.9\pm0.3$   \\
BD\,+59$\degr$385 & 6.76\,$\pm$\,0.06 &6.70\,$\pm$\,0.02 & 96.5\,$\pm$\,0.8     & 98.1\,$\pm$\,0.1  \\
Hiltner\,960 &   $5.70\pm0.13$ & $5.66\pm0.02$ & $56.3\pm0. 6$    &$54.5\pm0.2$   \\
Hiltner\,960 & $5.86\pm0.14$& $5.66\pm0.02$ & $53.5\pm0.3$   & $54.5\pm0.2$  \\
HD\,7927* &  $3.02\pm0.07$ & $3.30\pm0.03$  & $90.8\pm0.8$   & $91.1\pm0.2$    \\
HD 204827* & $5.15\pm0.14$ & $5.32\pm0.01$ & $57.7\pm0.8$     & $58.7\pm0.8$   \\
BD\,+64$\degr$106* & $5.18\pm0.10$ & $5.69\pm 0.04$  & $94.8\pm0.9$  & $96.6\pm0.2$ \\
\hline
\end{tabular}
\end{table*}

\begin{table*}[tb]

\vspace{3mm} \caption{A comparison of the measured parameters of
linear polarization for WOLL-2} \label{globdata:Afanasiev_n}
\medskip
\begin{tabular}{l|r|r|r|r}
\hline
\multicolumn{1}{c|}{Star}&\multicolumn{1}{c|}{$P_{\rm obs}$,\%}&\multicolumn{1}{c|}{$P_{\rm tab}$,\%}&\multicolumn{1}{c|}{$\rm PA_{obs}$, deg}& \multicolumn{1}{c}{$\rm PA_{tab}$, deg}\\
\hline
HD\,15597 & $4.50\pm0.27$ & $4.38\pm0.03$  & $102.9\pm1.7$   & $103.2\pm0.3$   \\
HD\,15597* & $4.13\pm0.33$ & $4.38\pm0.03$  & $100.1\pm2.6$     & $103.2\pm0.3$  \\
HD\,154445* & $3.14\pm0.24$ &$3.67\pm0.05$ & $84.1\pm5.1$     & $88.6\pm0.7$  \\
HD\,161056 & $4.02\pm0.19$ & $4.00\pm0.01$ & $66.2\pm1.4$    &$66.3\pm0.3$  \\
VI\,Cyg-12 & $8.97\pm0.07$ & $8.95\pm0.09$ & $115.2\pm1.2$   & $115.0\pm0.3$ \\
Hiltner\,960 &  $5.76\pm0.14$ & $5.66\pm0.02$  & $53.5\pm1.5$   & $54.8\pm0.2$   \\
HD\,204827* & $5.11\pm0.19$ & $5.34\pm0.02$ & $58.1\pm1.3$     & $58.7\pm0.8$  \\
\hline
\end{tabular}
\vspace{5mm}
\end{table*}

Figure~\ref{uncorr:Afanasiev_n} shows the effect of depolarization
by the Earth's atmosphere on the variation of the wavelength
dependence of the Stokes  $U$ and $Q$ parameters in the
S5\,0716+71 blazar with the synchrotron emission mechanism. It
should be noted that the spectra presented in
Fig.~\ref{uncorr:Afanasiev_n} are wavelength-modulated, which must
not be since the synchrotron radiation produces smooth wavelength
dependences of the Stokes parameters. If we interpret the
variation of depolarization during the observations (see
Fig.~\ref{atm:Afanasiev_n}) as a variation of the spectrograph
transmission for the ordinary and extraordinary rays, then in the
expression (13) instead of $\biggl(\displaystyle{\frac{I_{\rm
o}(\lambda)-I_{\rm e}(\lambda)}{I_{\rm o}(\lambda)+I_{\rm
e}(\lambda)}}\biggr)_k$ we should use
$\biggl(\displaystyle{\frac{D_{k}I_{\rm o}(\lambda)-I_{\rm
e}(\lambda)}{D_{k}I_{\rm o}(\lambda)+I_{\rm
e}(\lambda)}}\biggr)_k$, where $k$ is the number of the
corresponding exposure, and the depolarization coefficients $D_k$
are taken from the observed deviations relative to the mean
(Fig.~\ref{atm:Afanasiev_n}). The result of computation of the
Stokes parameters with the atmospheric depolarization corrected
this way is shown in Fig.~\ref{corr:Afanasiev_n}.

As shown in Fig.~\ref{corr:Afanasiev_n}, the result exceeds the
most audacious expectations---the modulation of the $U$ and $Q$
spectra is completely eliminated, while the statistical
measurement error has decreased 2.5-fold and corresponds to that,
expected for the Poisson statistics! This is a direct indication
that the discovered technique of correction for the depolarization
effects in the Earth's atmosphere caused by the scintillations is
correct. It should certainly be borne in mind that firstly, it is
applicable for the objects in which the times of internal
(intrinsic to the object) variability are longer than the total
exposure time, and secondly, this technique does not account for
the average degree of depolarization in the atmosphere. However,
it (the average degree of depolarization) can be determined from
the observations of the polarization standard stars before and
after the object at the same zenith distances.

\section{RESULTS OF TEST  OBSERVATIONS}

For the pilot observations of linear polarization we used the
polarization standard  stars from the lists of Hsu and
Breger~\citep{jin:Afanasiev_n}, and Turnshek et
al.~\citep{turn:Afanasiev_n}. The observations were conducted in
two polarization modes, WOLL-1 and WOLL-2. Whenever possible
during each night of observations,  the zero-polarization
standards were recorded   to control the instrumental
polarization, and the non-zero polarization standards were
registered to refine the zero-point of the position angle of the
polarization plane. The spectra were obtained in the range of
3700--8300 \AA\ with the spectral resolution of about 12~\AA\ with
the Volume Phase Holographic Grating VPHG940@600. After the series
of reductions, the sequence of which is described in the previous
section, the normalized Stokes vectors $Q(\lambda)$ and
$U(\lambda)$ were obtained for each star. The robust estimates of
the Stokes parameters in the photometric V-band were made taking
account of the standard V$(\lambda)$ curve. Next, using the (5)
and (15) relations, we calculated the average values of the
observed polarization  $P_{\rm obs}$  and the angle of the
polarization plane $\rm PA_{obs}$ in the photometric V-band, which
are listed in Table~1. The same table gives the true
values of  $P_{\rm tab}$ and $\rm PA_{tab}$, adopted
from~\citep{jin:Afanasiev_n} and~\citep{turn:Afanasiev_n}.

The estimates made during the night with unstable atmospheric
conditions (cirri) are marked with an asterisk in the table. The
estimates of polarization of standard stars demonstrated the
absence of  significant linear instrumental polarization within
the errors (of about 0.1\%). As we can see from the table, the
quality of data we have acquired varies greatly. On the nights
with good transparency the true accuracy of linear polarization
measurements reaches 0.05\% in the observations using the WOLL-1
polarization filter.

In cases where the transparency was not exceptional, and
depolarization reached 2--3\%, the observations were conducted
only with the double Wollaston prism (WOLL-2), which yielded
the measurement accuracy of about 0.2\%, what is well illustrated
by the data from Table~2, listing the measurements of the
polarization standards with WOLL-2.

\section{CONCLUSION}

The test observations and their analysis imply that the proposed
technique allows obtaining real values of polarization and making
sufficiently reliable estimates of their errors. Note in
particular that the proposed technique, given all the difficulties
of performing the polarimetric measurements, allows for the
measurements of linear and circular polarization in rather
difficult weather conditions. The observations of extended objects
in our case are not aggravated by the presence of the variable
instrumental polarization along the slit. This makes our technique
especially promising for the studies of faint starlike and
extended objects.

\begin{acknowledgements}
The authors are grateful to N.~V.~Borisov for help in purchasing
the polarization elements and for useful comments. This work was
supported by the Ministry of Education and Science of Russian
Federation (state contracts no.~14.740.11.0800, 16.552.11.7028,
16.518.11.7073) and the Russian Foundation for Basic Research
(grant no.~\mbox{12-02-00857-a}).
\end{acknowledgements}

 \end{document}